\documentclass[a4paper,fleqn]{cas-dc}
\usepackage{mathrsfs}
\usepackage{amsmath,amssymb,amsfonts}
\usepackage{algorithm}
\usepackage{algorithmicx}
\usepackage{algpseudocode}%

\usepackage{graphicx}
\usepackage{textcomp}
\usepackage{xcolor}
\usepackage{flushend}
\usepackage{url}
\usepackage{subcaption}

\usepackage{multirow}%
\usepackage{amsthm}%
\usepackage{mathrsfs}%
\usepackage[title]{appendix}%
\usepackage{manyfoot}%
\usepackage{booktabs}
\usepackage{listings}%

\usepackage[authoryear]{natbib}

\def\tsc#1{\csdef{#1}{\textsc{\lowercase{#1}}\xspace}}
\tsc{WGM}
\tsc{QE}

\begin{document}
\let\WriteBookmarks\relax
\def\floatpagepagefraction{1}
\def\textpagefraction{.001}

\shorttitle{Too many clients spoil the logs: distributed parsing, aggregation techniques and metrics}

\shortauthors{Garc\'\i a G\'omez, Beck et~al.}

\title[mode=title]{Too many clients spoil the logs: distributed parsing, aggregation techniques and metrics}


\author[1,2]{Andr\'e Garc\'\i a G\'omez}[orcid=0000-0003-3102-6986]
\ead{andre.garcia-gomez@ait.ac.at}

\author[1,3]{Viktor Beck}[orcid=0009-0003-0784-5479]
\ead{viktor.beck@ait.ac.at}

\author[1]{Max Landauer}[orcid=0000-0003-3813-3151]
\ead{max.landauer@ait.ac.at}

\author[1]{Markus Wurzenberger}[orcid=0000-0003-3259-6972]
\ead{markus.wurzenberger@ait.ac.at}

\author[1]{Florian Skopik}[orcid=0000-0002-1922-7892]
\ead{florian.skopik@ait.ac.at}

\author[3]{Andreas Rauber}[orcid=0000-0002-9272-6225]
\ead{rauber@ifs.tuwien.ac.at}

\author[2]{Edgar Weippl}[orcid=0000-0003-0665-6126]
\ead{edgar.weippl@univie.ac.at}

\affiliation[1]{organization={Cyber Security \& Communication Technologies, Austrian Institute of Technology},
  addressline={Giefinggasse 2},
  city={Vienna},
  postcode={1210},
  country={Austria}}

\affiliation[2]{organization={Computer Science, University of Vienna},
  addressline={Universit\"atsring 1},
  city={Vienna},
  postcode={1010},
  country={Austria}}

\affiliation[3]{organization={Information Systems Engineering, Technical University of Vienna},
  addressline={Karlsplatz 13},
  city={Vienna},
  postcode={1040},
  country={Austria}}



\begin{abstract}
Log messages are employed to identify anomalies in IoT systems and have been widely used across many applications. To enable collaboration among infrastructures belonging to different organizations, federated log anomaly detection has been thoroughly investigated for this purpose, often assuming the availability of ground truth log templates. However, this is typically not the case, as it is still necessary to depend on log parsers to extract log templates at inference time, while the design and deployment of such parsers in distributed environments remains largely an open problem. To this end, we report on an empirical investigation of distributed parsing and its key challenges, providing insights and lessons learned towards closing the gap between research and application. To further refine and contextualize the resulting benchmarks, we examine and introduce new metrics that provide a more comprehensive view for evaluating distributed parsers and how they behave in distributed environments. We propose a set of lessons learned and recommendations to support the design, deployment, and evaluation of distributed parsers. In particular, we evaluate the performance of five distinct parsing algorithms in combination with aggregation techniques that allow a unified agreement of event-template mappings across all participating clients, a prerequisite for federated log anomaly detection. In total, 6,720 distinct distributed configurations were tested for this purpose.
\end{abstract}




\maketitle

\section{Introduction}
\label{sec:introduction}

The earliest uses of logs date back to the introduction of syslog in Unix with 4.2BSD in 1983 \citep{stevens2013advanced}. Nearly half a century later, logging has become a fundamental component of system administration, as logs capture the different operational states of a system. In the current era, characterized by the extensive use of cloud computing and the Internet of Things (IoT), log processing has grown substantially more challenging due to the requirement to monitor logs produced concurrently by numerous machines participating in multiple services simultaneously \citep{raptisSurveyNetworkedData2023}. 

Although logs can serve many purposes, one of their most common uses is log anomaly detection. Because these techniques are used to pinpoint system failures and abnormal behaviors related to security breaches, they have emerged as a key research topic \citep{guo2021logbert, du2017deeplog}. A common requirement across most of these studies is the use of log parsing methods to transform raw logs into structured representations that are easier to analyze. With the growing number of IoT deployments in heterogeneous environments, several works have addressed the development of log anomaly detection solutions for distributed systems \citep{garciagomezCollaborativeAnomalyDetection2025}; however, comparatively little focus has been placed on designing and assessing distributed log parsing algorithms themselves, as most existing studies assume that logs have already been parsed before they are distributed across nodes. We refer to parser algorithms as the techniques that derive the templates used to handle the various logs. This paper investigates distributed parsing and examines its main challenges, including how parsing data is shared and managed across multiple clients (i.e., system nodes). We structure the paper around three central research questions (RQ).

\textbf{RQ1 How do the number of clients and the volume of log data influence the effectiveness of different parser algorithms?} We evaluate several parsing algorithms across diverse data distributions, datasets, and client counts. To further refine and contextualize the resulting benchmarks, we will additionally examine and discuss the metrics currently used in the literature and introduce new metrics that provide a more comprehensive view of the different parsers and how they behave in distributed environments. We define effectiveness as the extent to which the parser can generate templates that are equivalent to the ground-truth templates.

\textbf{RQ2 To what extend can different aggregation techniques improve the performance of local parsers in a distributed environment?} We define aggregation as the set of techniques used to transfer the knowledge required for parsing operations among the different local parsers. We evaluated a range of aggregation methods over several benchmarks to study the performance gains they can provide. To the best of our knowledge, no federated parsing strategy has been introduced in the existing literature. FedAvg \citep{mcmahan2017communication}, the most widely used aggregation method in deep learning, simply averages the model weights. Although this approach is not directly applicable to parsing, we adopt a similar philosophy by examining straightforward techniques such as set concatenation and merging strategies that have been proposed in prior work.

\textbf{RQ3 What are the main considerations when deploying parsers in distributed environments?} Based on the previous results, we will examine the key characteristics identified and propose a set of takeaways and recommendations to support the design and deployment of parsers in distributed environments.

The main contributions of this paper are summarized below:
\begin{itemize}
    \item A thorough examination of the various metrics found in the literature, along with the introduction of a taxonomy that categorizes these metrics and explains their relevance and application in distributed parser deployments.
    \item A comprehensive benchmarking study of multiple parsers operating in distributed environments, both with and without the use of different federated techniques.
    \item A set of lessons learned and recommendations regarding the deployment and use of parsers in federated environments.
\end{itemize}

We organize this paper into the following sections. Section \ref{sec:background} provides a brief overview of the different parsers and aggregation techniques. In Section \ref{sec:related_work}, we analyze related work found in the literature. Section \ref{sec:methodology} explains how parsers are deployed in distributed environments, outlines the techniques applied in this study, and presents an extensive analysis of the various metrics. Section \ref{sec:experiment_design} describes the empirical experiments. Section \ref{sec:results} presents the results of the empirical benchmarks. Section \ref{sec:disc} examines these results and derives recommendations for the development and deployment of distributed log parsers. In Section \ref{sec:validity}, we discuss threats to validity. Finally, Section \ref{sec:conclusion} summarizes our conclusions and outlines the directions for future work.

\section{Background}
\label{sec:background}
\begin{figure}
    \centering
    \includegraphics[width=\linewidth]{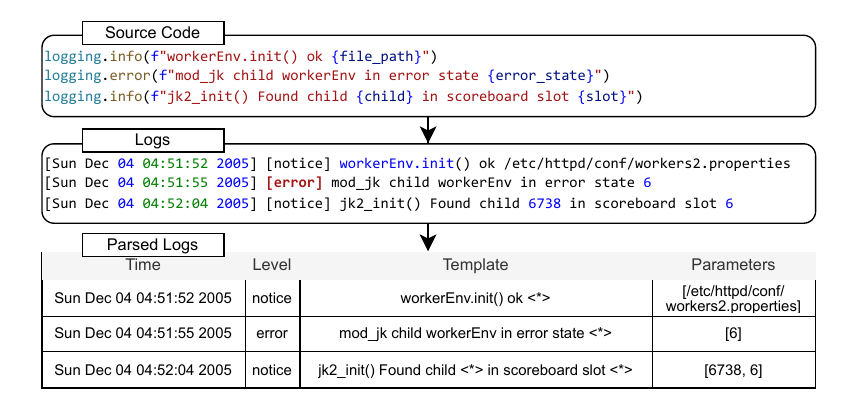}
    \caption{The log parsing process from the logging statement to a structured tabular representation \citep{beck2025system}.}
    \label{fig:parsing_process}
\end{figure}

Log parsing is the process of transforming logs into a structured format, which is required for downstream tasks like anomaly detection. The original logging statements from the source code are often not available. Therefore, algorithms have been proposed to extract these log templates directly from the logs. Once the template is known for a type of log, parsing becomes trivial and can be achieved by simple regular expression matching. Log parsing is therefore formulated as a template extraction problem \citep{heDrainOnlineLog2017}. An example of a typical log parsing process is given in Fig. \ref{fig:parsing_process}, where we see how the source code statements produce the logs and how they are then separated into headers (``Time'' and ``Level''), the template (defining the event type) and the parameters.  The template is always the same per event type and we therefore call it the static part of the log, whereas the parameters are variable.

\subsection{Types of Parsers}

The work of \citet{jiangLargeScaleEvaluationLog2024} identifies four different categories of parsers based on their parsing approach, namely frequency, similarity, heuristic, and semantic-based approaches. Since the general capabilities of LLMs also include the ability to evaluate the structure of logs beyond semantics, this list is extended to include LLM-based parsers \citep{beck2025system}.
\begin{itemize}
    \item \textbf{Frequency-based parsers}: Frequency-based parsers \citep{daiLogramEfficientLog2022, vaarandi2015logcluster} operate on the principle that tokens that appear frequently within a log dataset are likely to be part of the static log template. These parsers begin by traversing the log data to build item sets of frequent tokens. These frequent itemsets are then used to derive the corresponding log templates for the messages.

    \item \textbf{Similarity-based parsers}: Similarity-based parsers \citep{shimaLengthMattersClustering2016, hamooniLogMineFastPattern2016} treat log parsing as a clustering problem. They group logs messages into distinct clusters based on their similarity, with the assumption that all logs messages within a single cluster share the same template. Different methods within this category use various clustering algorithms, such as hierarchical or density-based clustering, and define similarity in different ways, using longest common subsequence (LCS), Jaccard index or likewise as measures of similarity. After clustering, the log template for each group is derived by extracting the common tokens from the logs in that cluster \citep{jiangLargeScaleEvaluationLog2024}.
    
    \item \textbf{Heuristic-based parsers}: Heuristic-based parsers \citep{heDrainOnlineLog2017, yuBrainLogParsing2023, jiangAbstractingExecutionLogs2008} employ a diverse range of heuristic algorithms and data structures to differentiate templates from parameters. These parsers are designed to take advantage of the unique characteristics of log data.
    
    \item \textbf{Semantic-based parsers}: Semantic-based parsers \citep{leLogParsingPromptBased2023, xuUniLogAutomaticLogging2024, huoSemParserSemanticParser2023} utilize neural networks to understand the semantic meaning of logs, to improve the parsing accuracy. These parsers are typically supervised methods employing bidirectional long short-term memory (LSTM) or pre-trained language models, mostly from the BERT family \citep{devlinBERTPretrainingDeep2019}. By training with labeled logs, they can distinguish between template components and parameters by treating it as a classification task.
    
    \item \textbf{LLM-based parsers}: LLM-based log parsers represent a new approach to automated log analysis and an extension to semantic-based parsers, leveraging Large Language Models (LLMs) to transform unstructured log messages into structured queryable data \citep{beck2025system}. These parsers offer advantages such as adaptability to various log formats and the ability to generalize across different system logs, which traditional methods often struggle with \citep{jiangLILACLogParsing2024}. Systems like LogBatcher \citep{xiaoDemonstrationFreeMorePractical2024} or LILAC \citep{jiangLILACLogParsing2024} integrate LLMs into multi-component frameworks to effectively use both syntactic and semantic information in logs. While some LLM-based parsers require labeled data for fine-tuning, others operate in an unsupervised manner. Techniques like in-context learning (ICL), retrieval-augmented generation (RAG), and caching are used to improve efficiency and accuracy.
\end{itemize}

\subsection{Parsing Caches}\label{sec:parsing-caches}
Another relevant concept for understanding the methodology of this work is the concept of a parsing cache. The term \textit{parsing cache} originates from the LLM-based parser LILAC \citep{jiangLILACLogParsing2024}. Parsing caches were initially designed to improve both the efficiency and consistency of LLM-based log parsing by storing previously generated log templates and reusing them when similar log messages appear. Since the number of unique log templates in real systems is significantly smaller than the number of log messages, caching parsed templates allows the system to avoid repeated LLM queries, and thus greatly reduce computational overhead. Although previous parsers such as Drain \citep{heDrainOnlineLog2017} have already implemented a tree-like structure for storing templates for efficient matching, parsing caches extend this functionality by adding specific matching and updating procedures. Thus, they can effectively handle the randomness in the templates that are sometimes generated by the LLM \citep{xiaoDemonstrationFreeMorePractical2024, jiangLILACLogParsing2024}.

LILAC's parsing cache is also implemented as a tree structure in which each log template is tokenized and stored as a path from the root to a leaf node. This structure allows efficient template matching and retrieval through tree traversal and also groups similar templates within the same subtree, which facilitates later refinement operations \citep{jiangLILACLogParsing2024}.

Cache matching is performed whenever a new log message is processed. The message is tokenized and sequentially matched against nodes in the cache tree. If traversal reaches a leaf node, the corresponding template is considered a successful match and is directly reused, avoiding an LLM query for template extration. If no leaf node is reached, we call it a cache miss and the traversal stops at the internal nodes and the templates in their subtrees are collected as relevant templates, which share partial prefixes with the input message \citep{jiangLILACLogParsing2024}. In such a case, the cache matching fails. The system invokes the LLM to generate a new template and performs cache updating. The generated template is compared with the relevant templates using a similarity measure based on the longest common subsequence of their token sequences. If the similarity is below a predefined threshold, the new template is inserted into the cache as a new entry. Otherwise, the existing template is refined by replacing differing tokens with a wildcard (usually represented by \texttt{<*>} in the literature \citep{khanGuidelinesAssessingAccuracy2022}), thus generalizing the template and ensuring consistent parsing across similar log messages \citep{jiangLILACLogParsing2024}.

\citet{xiaoDemonstrationFreeMorePractical2024} built upon LILAC's parsing cache and, additionally to each stored template in the cache, store a reference log to verify that logs that are grouped together in a preceding clustering process also match the same logs. When some logs cannot be matched, the template is considered valid for the subset of logs it successfully matches. The remaining unmatched logs are subsequently assigned to a new cluster, which is then placed back into the queue for further parsing.

\section{Related work}
\label{sec:related_work}

Log parsing has been extensively investigated in the literature. For instance, one of the oldest structured comparisons \citep{he2016evaluation} analyzes several parsers in combination with principal component analysis (PCA) as a standalone method for log anomaly detection. However, this work is now more than a decade old and evaluates parsers whose performance is no longer state of the art; in particular, the widely adopted parser Drain \citep{heDrainOnlineLog2017} was proposed only after that study appeared. To our knowledge, \citet{fu2023empirical} is the first work to explicitly raise concerns about the mismatch between log parsers and log anomaly metrics. That study also investigates how parsing errors propagate to the overall performance of anomaly detectors. It is important to note that, although the study examines a wide variety of log anomaly detectors, they all rely solely on log event IDs (representing event types) and ignore the parameters and syntax contained in the logs, which significantly limits the scope of the comparisons. The only exception is LogRobust \citep{zhang2019robust}, which additionally leverages log templates.

A similar conclusion is drawn by \citet{fu2023empirical} and later by \citet{khan2024impact}, who recommend the use of the distinguishability concept to tackle this issue. This concept employs minimal distinguishability as the criterion for characterizing perfect parsers, following the formalization of \citet{shin2021theoretical}, who propose an ideal definition of log parsing. Another important contribution in this area is that of \citet{le2022log}, who likewise study how log parsing errors influence log anomaly detection techniques. Their findings show that such errors can degrade performance, although detectors that make use of log templates tend to be more robust to these issues. Another study by \citet{sedkiAMLAccuracyMetric2024} investigates how the choice of different metrics in the literature influences the reported outcomes. \citet{beck2025system} assess the effectiveness of LLM-based parsers and identify the components that most strongly influence their performance, experimenting with 12 different parser algorithms, including both LLM-based and non-LLM approaches. Among these are parsing caches, which we also employ in this work.

To the best of our knowledge, one aspect that has not been explored is the use of federated parsing in distributed systems. This gap does not arise from a lack of interest or practical relevance, as there are several proposals for federated log anomaly detection methods that rely on such parsers, as illustrated by \citet{garciagomezCollaborativeAnomalyDetection2025}. However, most of these studies \citep{himler2024anomaly, nguyenDIoTFederatedSelflearning2019} assume that the data has already been parsed beforehand. Although this approach is sufficient to demonstrate the effectiveness of federated learning for log anomaly detection, it does not address how the data would actually be parsed in real-world deployments.

\section{Methodology}
\label{sec:methodology}
In this section, we introduce the overall pipeline and outline the scenarios in which federated parsing can be employed. We also describe several aggregation strategies for federated learning and analyze the different metrics that will be used in the empirical experiments.
\subsection{Federated pipeline}
\label{sec:pipeline}

Most distributed architectures rely on a collection of applications that operate independently, with their logs aggregated and stored through a publish–subscribe service \citep{raptisSurveyNetworkedData2023}. However, these logs are typically unstructured, which complicates their analysis. Parser algorithms are commonly used \citep{beck2025system} to transform them into structured data that can be used to detect faults or security threats in the system, or to support monitoring through dashboards.

\begin{figure*}
    
  \centering
  \begin{subfigure}[t]{0.49\linewidth}
    \centering
    \includegraphics[width=\linewidth]{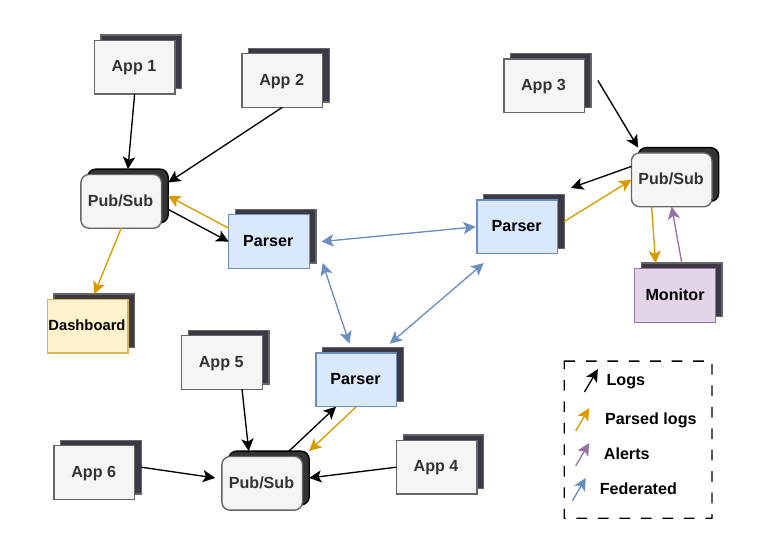}
    \caption{Federated pipeline across multiple distributed systems.}
    \label{fig:federated-pipeline}
  \end{subfigure}\hfill
  \begin{subfigure}[t]{0.49\linewidth}
    \centering
    \includegraphics[width=\linewidth]{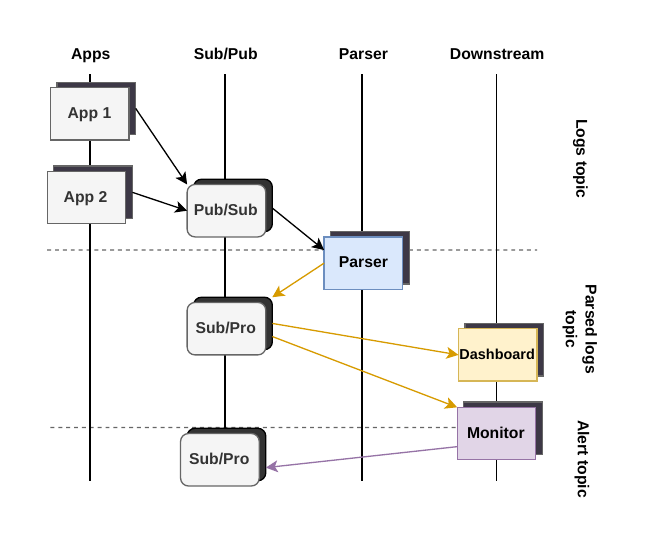}
    \caption{Distributed logging communication in a streaming pipeline.}
    \label{fig:stream-pipeline}
  \end{subfigure}
  \caption{Pipeline diagrams used throughout this section.}
  \label{fig:pipeline-diagrams}
\end{figure*}

However, widely used parsing algorithms, such as Drain \citep{heDrainOnlineLog2017}, must be trained on the data and need to encounter each type of log at least once to achieve optimal performance. This poses challenges when systems are constantly evolving or when specific applications or services are present on some nodes but almost never appear on others. Federated parsing can alleviate this problem by enabling parsers from different organizations to be jointly utilized in the training process. This is illustrated in Figure \ref{fig:federated-pipeline}, which presents an example where multiple service organizations each operate their own parser within a microservice architecture \citep{raptisSurveyNetworkedData2023}, and demonstrates how logs flow from the various applications to the parsers and subsequently from the parsers to downstream tasks in a federated parsing environment.

A commonly overlooked requirement for parsers is that they must operate in a streaming fashion. This means that for every log message they receive, they must output a parsed log close to real time, ensuring that the pipeline remains intact, as illustrated in Figure \ref{fig:stream-pipeline}. Although most parsers can update their templates in real time, this capability can cause issues for downstream tasks, since the revised templates may also alter how previously seen log types are interpreted. In this work, we will also consider this requirement.

\begin{figure*}[!t]
  \centering
  \includegraphics[width=0.7\textwidth]{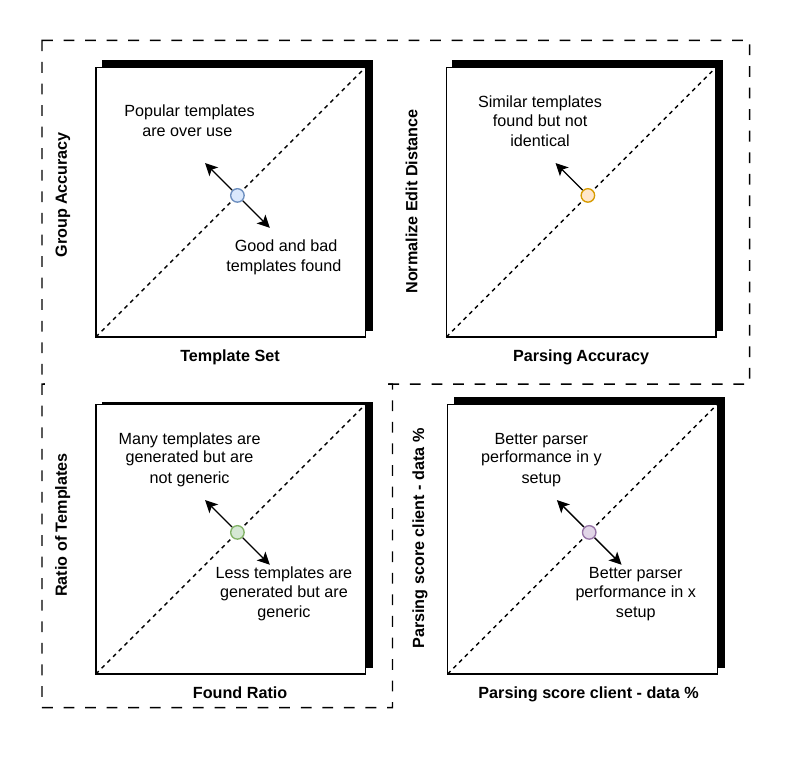}
  \caption{Explanation of the different result plots.}
  \label{fig:plot_comparison}
\end{figure*}

\subsection{Parser definition}

To understand what a parser actually does, we first need to define what a log is. Logs are generated by log statements ($\mathcal{S}_c$), where $c$ denotes the constant message of the log. Each time the log statement is executed, it receives as input a set of variables ($\mathcal{V}$), which are used to produce the final log message ($l$), formally represented as $\mathbf{S_c}(\mathcal{V}) \rightarrow l$. These variables can be any data describing the current state of execution, such as timestamp, login details, and so on. For example, the log $l=$\texttt{Hello Thomas, you have successfully logged in to App 4.} can be decomposed into a constant part $c=$\texttt{Hello <*>, you have successfully logged in to <*>}. and a set of variables $\mathcal{V} = \{\text{\texttt{Thomas}}, \text{\texttt{App 4}}\}$. The variable components of the event type are crucial for understanding the current execution state and for detecting anomalous behavior related to faulty execution or system intrusions. Nevertheless, isolating the variables from the constant segment is challenging because in many practical cases when the source code is not available, the constant part is not known in advance.

\subsubsection{Mathematical definition}
As discussed earlier, parsing algorithms are techniques that transform an unstructured log ($l \in L$) into a structured log ($\hat{l} \in \hat{L}$). They achieve this by constructing a set of templates $\mathcal{T}$, where each template corresponds to the inferred constant portion of a log message produced by the parser. Formally, we define a parser as $Parser(l, \mathcal{T}_i) \rightarrow \hat{l}, \mathcal{T}_{i+1}$.

If a new log type appears and updates to the parser are allowed, then we anticipate $\mathcal{T}_i \neq \mathcal{T}_{i+1}$, because the template set will be adjusted accordingly. In contrast, if the parser is fixed (for instance, to avoid impacting downstream processes), the introduction of a new log type will leave the template set unchanged ($\mathcal{T}_i = \mathcal{T}_{i+1}$). Under these conditions, the resulting parsed log $\hat{l}$ will either be parsed incorrectly or, if no template matches, the parser will output an empty parsed log $\hat{l}_{\varnothing}$.

Assuming a ground-truth template set $\mathcal{T}^*$, a good parser should yield $\mathcal{T}_i \approx \mathcal{T}^*$ when our log set $\mathcal{L} = \{ l_i \mid \forall i \in N \}$ is sufficiently representative of the overall distribution. The key idea behind using a federated approach is that, if $\mathcal{L}$ is not representative enough, aggregating the template sets from each node will result in a template set that is closer to the ground truth, without requiring all raw logs to be shared across nodes.

\subsubsection{Federated parsing}
In Pseudocode \ref{tab:federated-pipeline-alg}, we present the federated parser method. In each round, a batch of logs is collected and processed by the local parsers to produce new templates. These local templates are then merged into a single global template set, which is subsequently used to update all local templates. In realistic situations, the communication between parsers can be done in two ways:

\begin{itemize}
    \item \textbf{Centralized}: Whenever a parser discovers a new template, it forwards this template to a central server. The central server incorporates the new template into its collection and then distributes the updated set of templates to all clients. This follows a similar strategy to the one used in FedAvg \citep{mcmahan2017communication} (Section \ref{sec:related_work}).
    \item \textbf{Decentralized}: A Paxos-based protocol \citep{lamportPaxosMadeSimple2001} can be employed to enable communication among parsers without relying on a central server. Each time a parser obtains a new template, it initiates a voting process to determine the resulting global template set.
\end{itemize}

A straightforward approach would be to simply merge all sets into a single set without requiring any coordination among clients. However, for the parser to be applicable to anomaly detection methods such as Deeplog \citep{du2017deeplog}, each template must be assigned a unique Event ID. Otherwise, different clients will assign different Event IDs to the same template, making it difficult to later perform federated learning for anomaly detection, as done by \citet{himler2024anomaly}.

\begin{algorithm}
  \caption{Federated parsing pipeline}
  \label{tab:federated-pipeline-alg}
  \centering
  \begin{algorithmic}[1]
    \State\textbf{Input:} parsers $\mathcal{P}$, local logs $\{L_p\}_{p\in\mathcal{P}}$, rounds $T$
    \State\textbf{Output:} Log Templates $\mathcal{T}$
    \State $\theta \leftarrow \text{InitModel}()$
    \For{$t \leftarrow 1$ \,to\, $T$}
      \ForAll{parsers $p \in \mathcal{P}$ \textbf{in parallel}}
        \State$L_p \leftarrow \text{GatherNewLogs}()$
        \State$\mathcal{T}_p, \_ \leftarrow \text{Parser}(\mathcal{T}_p, L_p)$
      \EndFor
      \State$\theta \leftarrow \text{Aggregate}(\{u_c\}_{c\in\mathcal{C}})$
      \ForAll{parsers $p \in \mathcal{P}$ \textbf{in parallel}}
        \State$\mathcal{T}_p \leftarrow\text{UpdateLocalTemplates}(\mathcal{T})$
      \EndFor
    \EndFor
  \end{algorithmic}
\end{algorithm}



\subsection{Aggregation strategies}
Existing works about federateed log anomaly detection often assume the templates to already be available, thus working with the groundtruth templates of the commonly used log datasets, such as the LogHub datasets \citep{zhuLoghubLargeCollection2023}. Many of these federated methods only work under the assumption that there is a unified agreement of which event is mapped to which Event ID. For instance, \citet{himler2024anomaly} learn event ID sequences. Other works, such as the detectors described by \citet{landauer2023aminer}, use the variables of specific events for detection. However, this is also only possible when the events are clearly defined since the variable of an event $X$ may have absolutely no relation to an event $Y$, thus a clear mapping is necessary for the algorithms to learn the relevant aspects of the data. To this end, we propose aggregation techniques that allow the creation of a unified agreement of which template is which event (or Event ID) across all participating clients.

\subsubsection{Concatenation}
Concatenation is the most straightforward aggregation method. It simply merges all unique templates into a single set and then assigns them the corresponding Event IDs.

\subsubsection{Parsing Caches}
As described in subsection \ref{sec:parsing-caches}, the original purpose of a parsing cache is to store extracted templates, merge similar templates and prune incorrect ones, and match logs to the corresponding templates. If a log is not able to be matched to an existing template, a template extraction process is triggered which can be completely separated from the parsing cache itself. Accordingly, this allows to produce the templates distributed and send them to a centralized parsing cache, or each clients own parsing cache. For our experiments we use the parsing caches of the log parsers LILAC \citep{jiangLILACLogParsing2024} and LogBatcher \citep{xiaoDemonstrationFreeMorePractical2024}.

\subsection{Metrics analysis}

\begin{figure*}[t]
  \centering
  \includegraphics[width=0.6\linewidth]{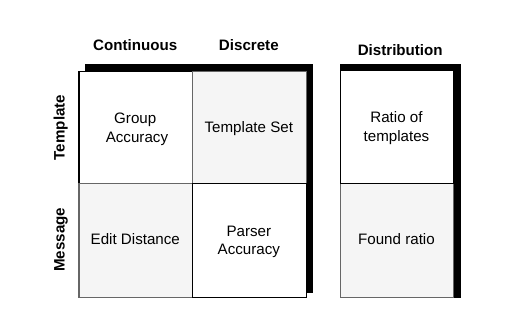}
  \caption{Table of the different metric groups.}
  \label{fig:table_metrics}
\end{figure*}

Although the literature offers many different types of parsing metrics, in this work, we do not simply report their values; instead, we provide a more detailed analysis of what these metrics represent and how they are applied to our results. Consequently, we observe that the various metrics can be grouped into the categories presented in Figure \ref{fig:table_metrics}.  This categorization follows the approach of \citet{sedkiAMLAccuracyMetric2024}, in which metrics are grouped into different types:

\begin{itemize}
    \item \textbf{Messages}: Metrics that verify whether the log message has been parsed correctly.
    \item \textbf{Template}: Metrics that assess whether the templates have been correctly identified.
    \item \textbf{Distribution}: Metrics that assess both the overall quality and the volume of the templates generated. Here, we define quality as the number of templates that, once created, can be used to match multiple logs. This set will be considered as an independent group, separate from the others.
\end{itemize}

We additionally classified the metrics into continuous and discrete groups according to their type. The various metrics we adopted from the literature are:

\begin{itemize}
    \item \textbf{Group accuracy (GA)} \citep{zhuToolsBenchmarksAutomated2019}: this metric only evaluates if the logs from the same template are assigned to the same group. It is formulated as:
    \begin{equation}
        GA = \frac{\sum_{s \in|\mathcal{I}(L^*) - \mathcal{I}(\hat{L})|} |s|}{|\hat{L}|}
    \end{equation}
    where $L^*$ denotes the ground-truth of the parsed logs, and $\mathcal{I}(\hat{L}) \rightarrow \{\{i \mid \forall \hat{l}_i.t = t\} \forall t \in \hat{L}.t\}$, with $\hat{l}.t$ indicating the template and $i$ the index associated with the parsed log message $l_i'$. We classify this metric as template-continuous because it relies on the template assignment to a group of events in a continuous way.
    \item \textbf{Parsing accuracy (PA)}: We follow the definition of \citet{daiLogramEfficientLog2022}, and define the metric as:
    \begin{equation}
        PA = \frac{\sum_{\forall i \in \hat{L}} l_i'.t = l^*_i.t}{|\hat{L}|}
    \end{equation}
    It computes how often we correctly assign the exact ground truth to the various log messages. This metric is classified as message-discrete, since it is only concerned with the specific template assigned to each individual message in a discrete manner.
    \item \textbf{Edit Distance (ED)}: For each message it measures how similar the ground-truth template is to the predicted template by computing the minimum number of operations (insertions, deletions, or substitutions) needed to convert one string into the other \citep{nedelkoskiSelfsupervisedLogParsing2021}. To align with the other metrics and keep the values within the [0, 1] range, we use the \textbf{normalized Edit Distance (NED)}:
    \begin{equation}
        NED = \frac{n - ED(l_i'.t, l^*_i.t)}{n}
    \end{equation}
    where $n = max(len(l_i'.t), len(l_i^*.t))$. We classified it as message-continuous.
\end{itemize}

During our analysis, however, we noticed that our taxonomy lacked a discrete template-based metric. To address this gap, we introduce a new metric, which we call \textbf{Template Set (TS)}. This metric evaluates whether the original templates are present in the parsed logs.

\begin{equation}
    \text{Template Set} = \frac{|\mathcal{T^*} \cap \mathcal{T}|}{|\mathcal{T^*}|}
\end{equation}
where $\mathcal{T}$ denotes the set of templates generated by the parser and $\mathcal{T^*}$ is the ground-truth template set. In this study, we excluded Precision and Recall template accuracy, as they do not provide additional insights \citep{khanGuidelinesAssessingAccuracy2022}. Finally, since the different distributions between nodes will produce different numbers of templates with different qualities, we propose two new metrics:

\begin{itemize}
    \item \textbf{Ratio of templates (RT)}: measures how many templates were actually produced compared to the total number of templates that were expected. This metric is independent of how templates are assigned.
    \begin{equation}
        RT = \frac{|\mathcal{T}|}{|\mathcal{T^*}|}
    \end{equation}
    \item \textbf{Found ratio (FR)}: determine how many logs were assigned a template; this metric does not consider whether the template assignment is correct or not.
    \begin{equation}
        FR = \frac{\sum_{\forall i \in \hat{L}} \hat{l}_i \neq \hat{l}_\varnothing}{|\hat{L}|}
    \end{equation}
\end{itemize}

\citet{sedkiAMLAccuracyMetric2024} proposed a new evaluation metric that combines both message-level and template-level scores. Their metric adapts an approach originally developed to assess satellite data accuracy. While the underlying idea is novel, we argue that their metric is not clearly consistent with standard practices in the parsing community, because it is borrowed from a completely different field with a distinct data domain and ignores well-established metrics that are routinely used for log processing. Therefore, we introduce a similar, but more conventional, metric that achieves the same goal by combining the previously presented metrics in a composite metric, we call it, \textbf{Parsing Score (PS)}:

\begin{equation}
    \label{eq:alpha}
   PS = \alpha TS + \beta NED + \gamma FR
\end{equation}

where $\alpha + \beta + \gamma \simeq 1$. This metric is defined as a weighted combination of three metrics from the previous categories, offering a more informative representation.


\section{Experiment design}
\label{sec:experiment_design}

In this section, we present the experimental setup along with all of its components. The code used to conduct these experiments is open source \footnote{\url{https://anonymous.4open.science/r/log_parsing_benchmark-49E4/README.md}; accessed 24-June-2026}.

\subsection{Experiment setup}
\label{sec:exp_setup}
To run the experiments, we developed a pipeline to gather all the results. For each parser–dataset combination, our goal was to evaluate how the amount of training data influences the performance of the parser and how splitting this data among multiple clients can help when only a small dataset is available. We also emulated a streaming scenario by training the parser on only a subset of the data while still using the entire dataset at prediction time. In doing so, we target the use case described in Subsection \ref{sec:pipeline}, where templates are fixed after training so as not to interfere with downstream tasks.

Other work in the federated log domain, such as that of \citet{himler2024anomaly}, not only distributes data evenly across clients to achieve an independent and identically distributed (IID) setup, but also evaluates performance under non-IID conditions to demonstrate that their methods remain effective when data are imbalanced. This is particularly important in deep learning, as these approaches typically rely on aggregation schemes such as averaging local models \citep{mcmahan2017communication}. In contrast, since we employ techniques similar to those of \citet{pustozerovaLightweightTechniquesFederated}, where no performance differences were observed between IID and non-IID settings, we restrict our study to an IID configuration to simplify the overall experiments.

\subsection{Data analysis}
We evaluate a total of 6,720 distributed parsing setups; each setup included several metrics, resulting in 40,320 data points to be analyzed. All metrics were logged and released as part of a public dataset inside the code repository, since it is not possible to display all results within this paper. To aid interpretation, we present a set of high-level plots that highlight the principal trends. Figure \ref{fig:plot_comparison} shows the different high-level plots and illustrates the expected behavior for the various outcomes. These plots compare the continuous and discrete versions of the same metric types defined in Figure \ref{fig:table_metrics}, with the exception of the final plot, where we use the same metric to compare aggregation techniques across different data distributions. 

If a data point in Figure \ref{fig:plot_comparison} lies on the dashed line, it indicates that both metrics are equivalent. However, if the results fall to one side or the other, this can be due to several factors. Below, we describe the first three plots from left to right:

\begin{itemize}
    \item \textbf{GA vs TS}: If TS increases but GA decreases, it means the parser was able to match more ground truth templates. However, it also discovered additional, more specific templates, which split some log type groups into smaller subgroups. If vise versa, it indicates that log types are being correctly grouped, but the templates assigned to those groups are not the ground truth templates.
    \item \textbf{NED vs PA}: If both metrics take the same value, it means that every correct log type has been matched with the right template. If NED increases while PA decreases, it indicates that the identified templates are similar to, but not exactly the same as, the ground-truth templates. The opposite situation cannot happen, because it would imply that generating the same template as the ground truth somehow decreases the similarity between templates.
    \item \textbf{RT vs FR}: If RT increases while FR decreases, it indicates that more templates are being generated; however, these templates are not representative of the overall dataset because they cannot be matched with future logs. Conversely, if RT decreases and FR increases, it means the generated templates are more generic.
\end{itemize}

\subsection{Parsers used}
Building on the work of \citet{jiangLargeScaleEvaluationLog2024}, four main categories of parsers are distinguished according to their parsing strategy: frequency-based, similarity-based, heuristic-based, and semantic-based approaches. Given that modern LLMs can analyze log structures beyond their semantic content, this categorization is expanded to include LLM-based parsers as an additional class \citep{beck2025system}. The parser types and approaches used in this study are described in Table \ref{tab:parsers}. We did not consider semantic-based parsers for two reasons: (1) They are supervised and require a significant number of templates for training, making it difficult to compare against other parsers that do not need templates --- especially, since distributed operation in parsing or anomaly detection is mostly practical for data scarcity on the individual clients \citep{leLogParsingPromptBased2023, liuUniParserUnifiedLog2022}. (2) The two most prominent examples LogPPT \citep{leLogParsingPromptBased2023} and UniParser \citep{liuUniParserUnifiedLog2022} are consistently outperformed in recent benchmarks against various LLM-parsers \citep{jiangLILACLogParsing2024, songParseLLMPriorFreeLLM2025, huangNoMoreLabelled2025}.
\begin{table}
    \caption{Parsers selection table}
    \centering
    \begin{tabular}{c|c}
         \textbf{Parser type} &  \textbf{Parsers}   \\\hline \hline 
         \textbf{Frequency-based} & Logram \citep{daiLogramEfficientLog2022}  \\
         \textbf{Similarity-based} & LenMa \citep{shimaLengthMattersClustering2016}  \\
         \textbf{Heuristic-based} & Drain \citep{heDrainOnlineLog2017}, Brain \citep{yuBrainLogParsing2023} \\
         \textbf{Semantic-based} & -  \\
         \textbf{LLM-based} & LogBatcher \citep{xiaoDemonstrationFreeMorePractical2024} \\
    \end{tabular}
    \label{tab:parsers}
\end{table}

\subsection{Dataset selection}

In our experiments, we selected HDFS~\citep{xuMiningConsoleLogs} and BGL~\citep{oliner2007supercomputers}, because they are the most widely used datasets in previous work \citep{garciagomezCollaborativeAnomalyDetection2025}. The datasets can be found on Loghub \citep{zhuLoghubLargeCollection2023}. Since each parser applies its own strategy to align logs with their final templates, which can introduce additional noise, we adopted the tree matcher method from DetectMate\footnote{\url{https://github.com/ait-detectmate/DetectMatePerformance}; accessed 03-June-2026} to standardize this process, DetectMate is an open-source software developed for log anomaly detection. This approach was used for every template-matching operation reported in the paper.

\begin{figure}
  \centering
  \includegraphics[width=0.8\linewidth]{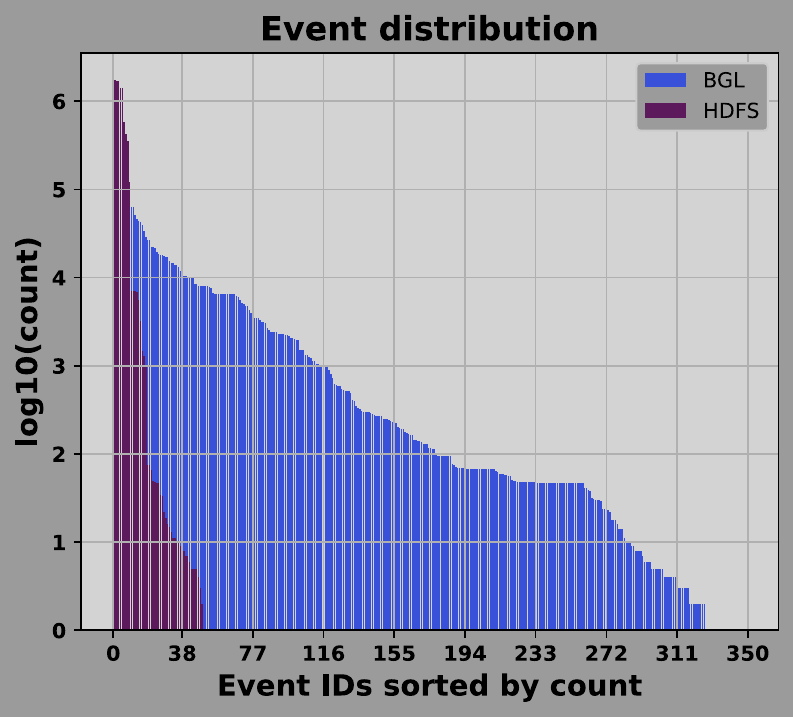}
  \caption{Event ID frequency distribution across the various datasets, ordered by number of occurrences.}
  \label{fig:dataset_dist}
\end{figure}

Table \ref{tab:datasets} lists the datasets used in this paper; the corresponding ground-truth templates can be found in our code \footnote{\url{https://anonymous.4open.science/r/log_parsing_benchmark-49E4/README.md}; accessed 24-June-2026}. Although DetectMate is fast and generally effective at matching templates, it failed to match 6\% of the BGL templates. This is partly because a small fraction of the logs do not have a correct template and because the ground truth does not define a generic template such as "$\texttt{<*>}$". By default, DetectMate removes all non-alphanumeric symbols, thereby filtering out noise that could otherwise interfere with the matching process, this technique has also been shown to enhance performance in prior work \citep{zhuMLParserEfficientAccurate2022}. The distribution of event types across both datasets is shown in Figure \ref{fig:dataset_dist}.
\begin{table}
    \caption{Datasets selection table}
    \centering
    \begin{tabular}{c|c|c|c}
         \textbf{Dataset} &  \textbf{Num. logs} & \textbf{Num. templates} & \textbf{FR}   \\\hline \hline 
         \textbf{HDFS} & 11,175,629 & 52 & 1.00  \\
         \textbf{BGL} & 4,713,493 & 377 & 0.94  \\
    \end{tabular}
    \label{tab:datasets}
\end{table}

\subsection{Setup}

All experiments were conducted on a laptop equipped with 16 GB of RAM, an Intel Core Ultra 7 165Ux14 CPU, and Ubuntu 24.04.3 LTS. The LLM-based parser LogBatcher \citep{xiaoDemonstrationFreeMorePractical2024} was run with \texttt{gpt-5.4-mini}\footnote{\url{https://developers.openai.com/api/docs/models/gpt-5.4-mini}; accessed 03-June-2026} of the latest model family of OpenAI, designed for high-volume workloads.

\section{Results}
\label{sec:results}

In this section, we report the performance of the different parsers in the distributed setting. We begin by examining the behavior of individual local nodes and then compare these outcomes with those obtained under a distributed configuration. A more detailed discussion of the results is provided in Section \ref{sec:disc}.

\subsection{Local performance}

For this experimental setup, we follow the pipeline illustrated in Figure \ref{fig:federated-pipeline}, where the data is streamed to a designated parser. This parser may use a fixed number of logs to finetune itself and must then parse the remaining logs without modifying the resulting templates. To evaluate this, we perform multiple runs in which we train the parsers with varying numbers of nodes and different amounts of training data per node. Each configuration is executed three times with randomly split data, and we report the average across all runs. Since each run contains a different number of nodes, we consistently use the entire dataset as test data on node 0. While one could argue that data previously seen by the parser should not appear in the test dataset, this is a standard practice in the log-parsing literature \citep{heDrainOnlineLog2017}. The parser is treated as a preprocessing component that incrementally adapts itself on the fly, rather than as a decision-support algorithm such as an anomaly detection system.

\begin{figure*}[t]
  \centering
  \includegraphics[width=0.75\textwidth]{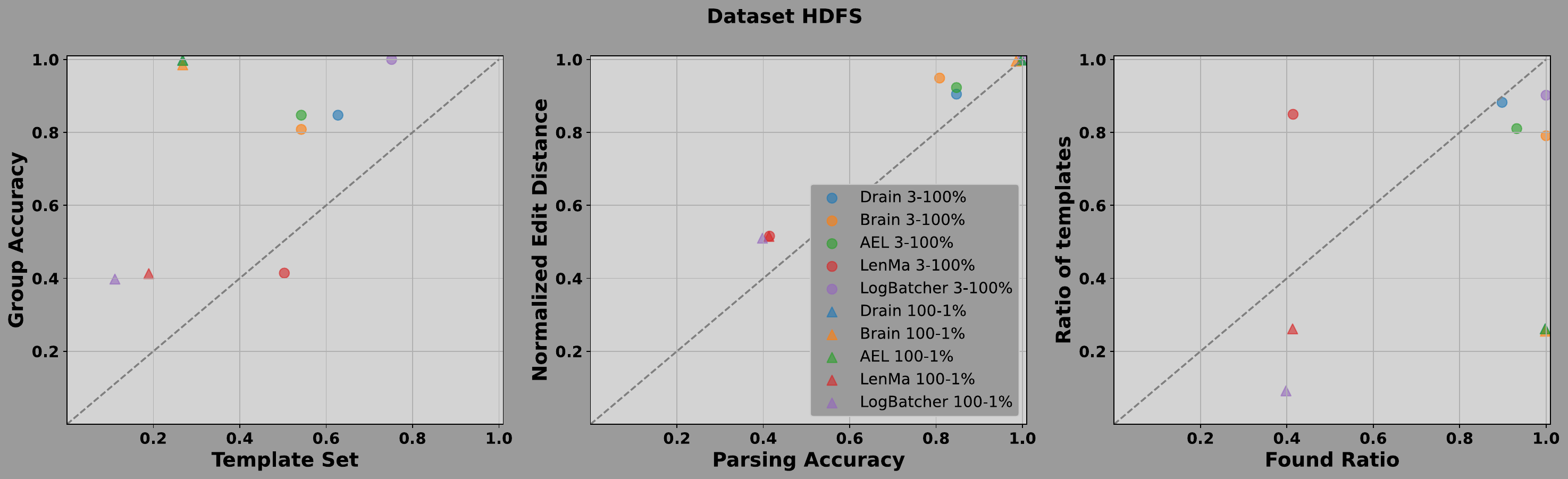}
  \caption{Plot comparison of the different metrics for the HDFS dataset.}
  \label{fig:hdfs-comparison}
\end{figure*}

\begin{figure*}[t]
  \centering
  \includegraphics[width=0.75\textwidth]{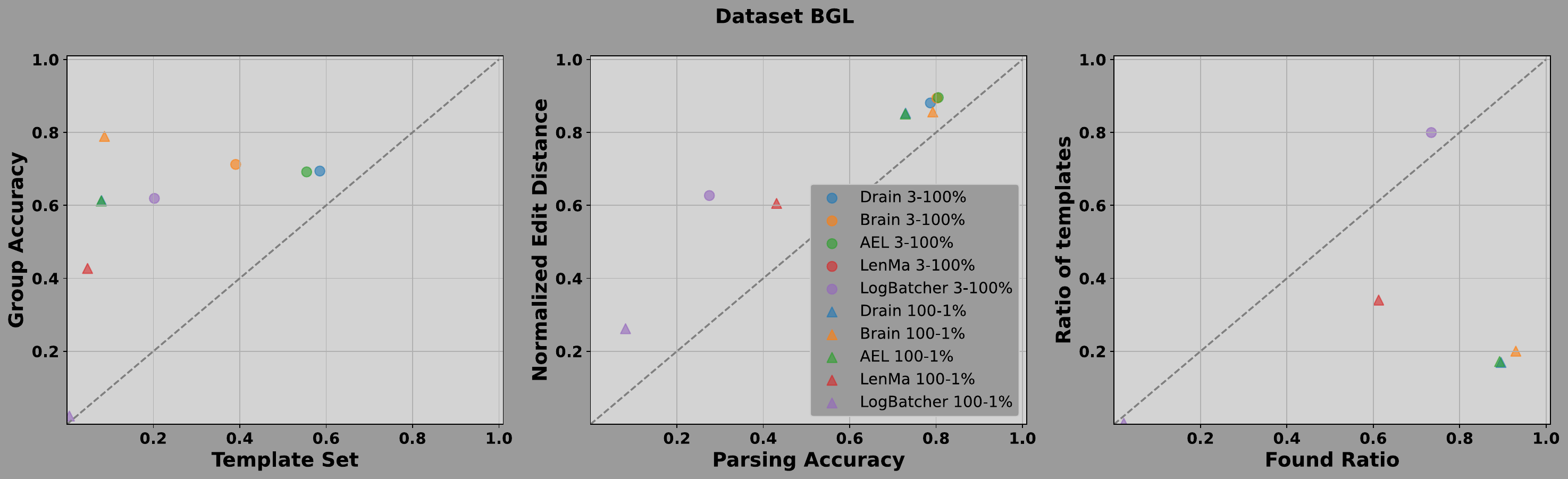}
  \caption{Plot comparison of the different metrics for the BGL dataset.}
  \label{fig:bgl-comparison}
\end{figure*}

Figures \ref{fig:hdfs-comparison} and \ref{fig:bgl-comparison} present the two most contrasting configurations examined in our experiments. At one end of the spectrum, we consider a simple three-node setup in which all streaming data is available for training. At the other end, we use 100 clients but restrict training to just 1\% of the data. The rationale behind these different plots is detailed in subsection \ref{sec:exp_setup} and visualized in Figure \ref{fig:plot_comparison}. Two clear trends emerge: certain parsers, such as LogBatcher, show reduced performance when both the number of clients and the data volume decrease. Conversely, other parsers, including Drain and Brain, are more resilient on smaller datasets and, in some cases, even perform better when trained with less data.

\begin{figure*}[t]
  \centering
  \includegraphics[width=0.6\textwidth]{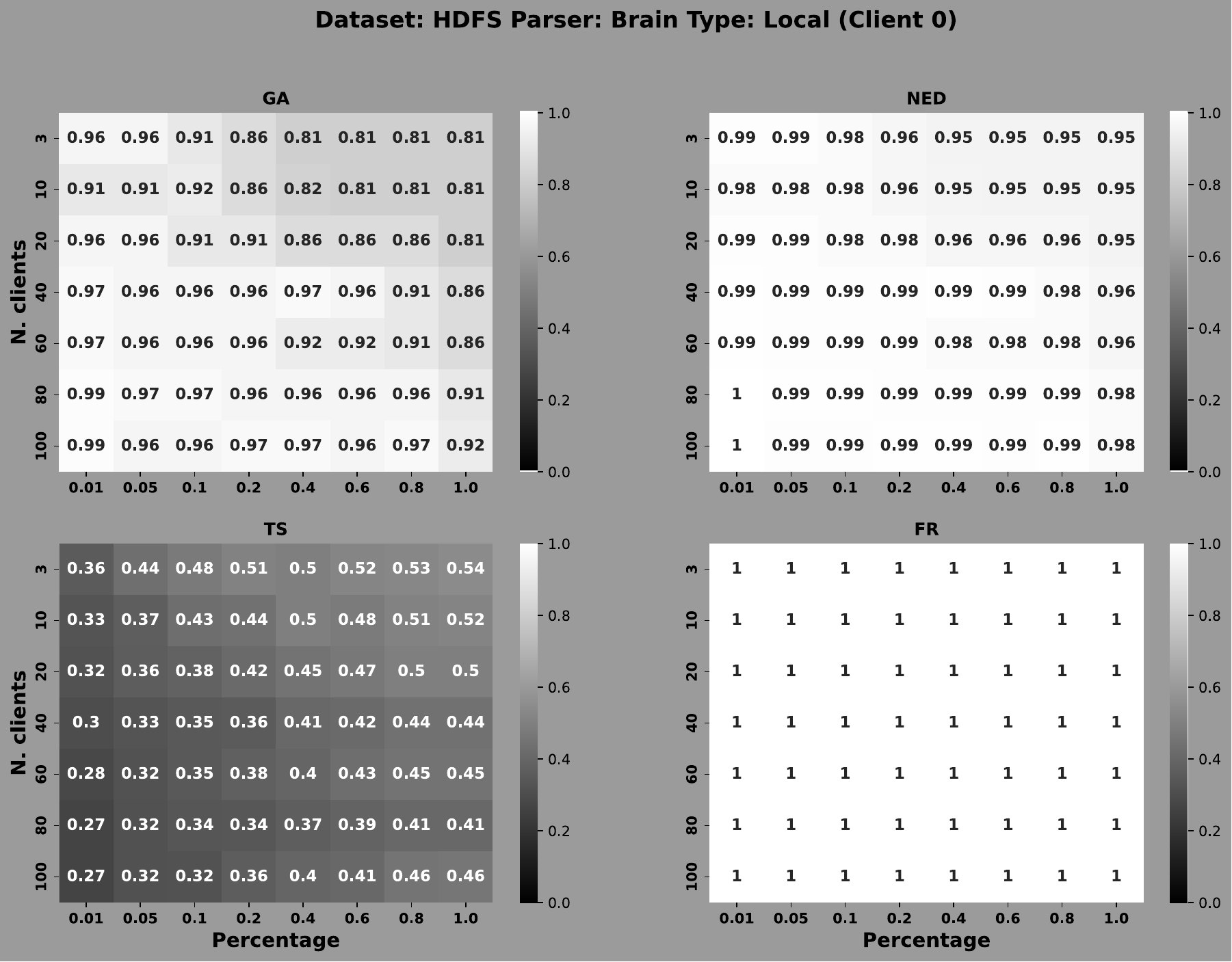}
  \caption{Heatmap illustrating the relationship between the number of clients and the proportion of data in a local configuration for the HDFS dataset with Brain.}
  \label{fig:hdfs-local_heat_map}
\end{figure*}

\begin{figure*}[t]
  \centering
  \includegraphics[width=0.6\textwidth]{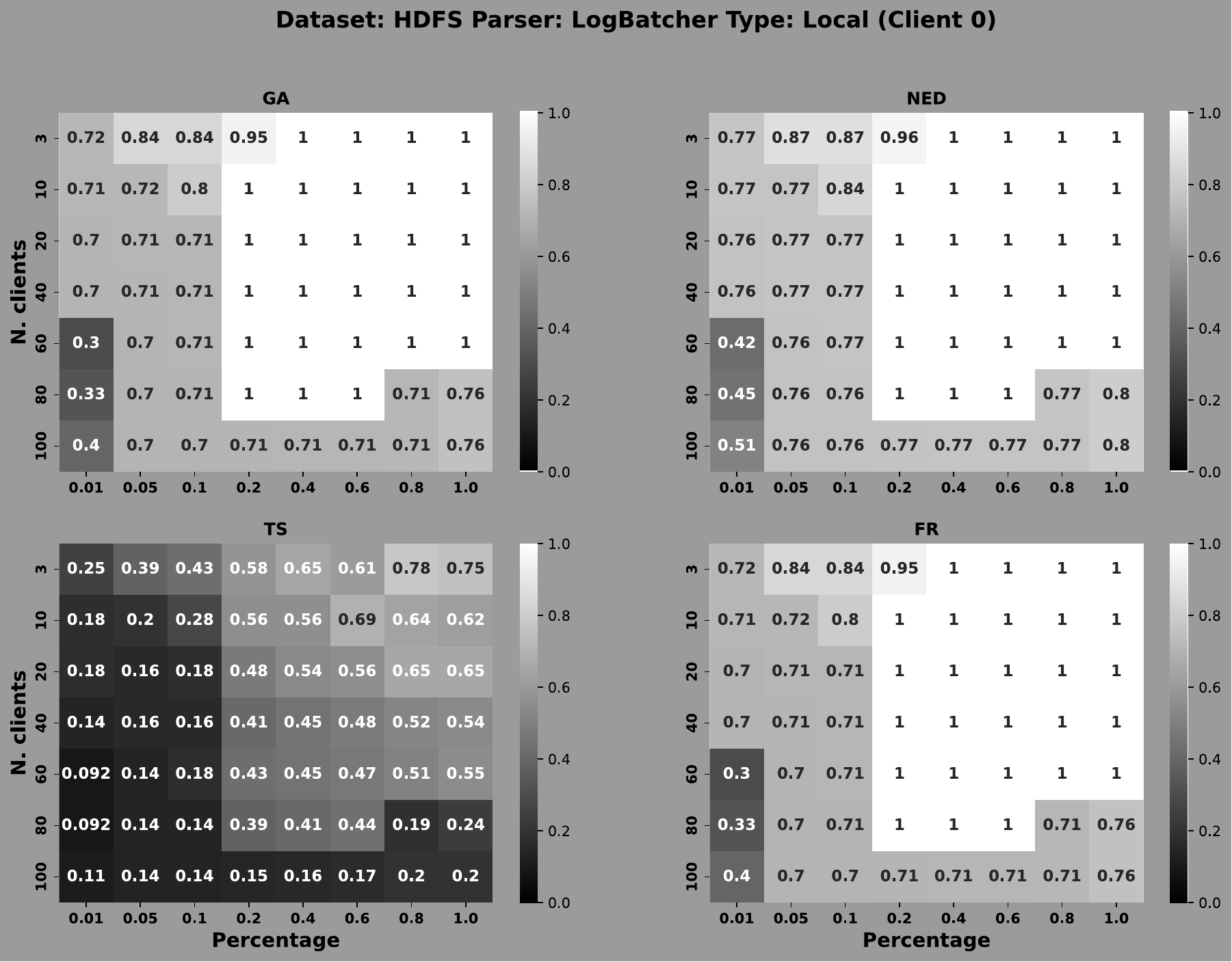}
  \caption{Heatmap illustrating the relationship between the number of clients and the proportion of data in a local configuration for the HDFS dataset with LogBatcher.}
  \label{fig:hdfs-local_heat_map_llm}
\end{figure*}

Figures \ref{fig:hdfs-local_heat_map} and \ref{fig:hdfs-local_heat_map_llm} show how different metrics vary with the degree of data distribution for Brain. Interestingly, in the HDFS scenario with only 3 clients, we observe that having more data does not always yield better performance. A similar phenomenon is reported by \citet{pustozerovaLightweightTechniquesFederated}, who note that, in a federated HDFS setting, increasing the data volume did not necessarily enhance anomaly detection performance. In line with their reasoning, we hypothesize that there exists a small subset of logs which, when included during training, can actually harm performance. In contrast, LogBatcher exhibits a more predictable performance gain as we reduce the number of clients and enlarge the dataset. 

\begin{figure*}[t]
  \centering
  \includegraphics[width=0.6\textwidth]{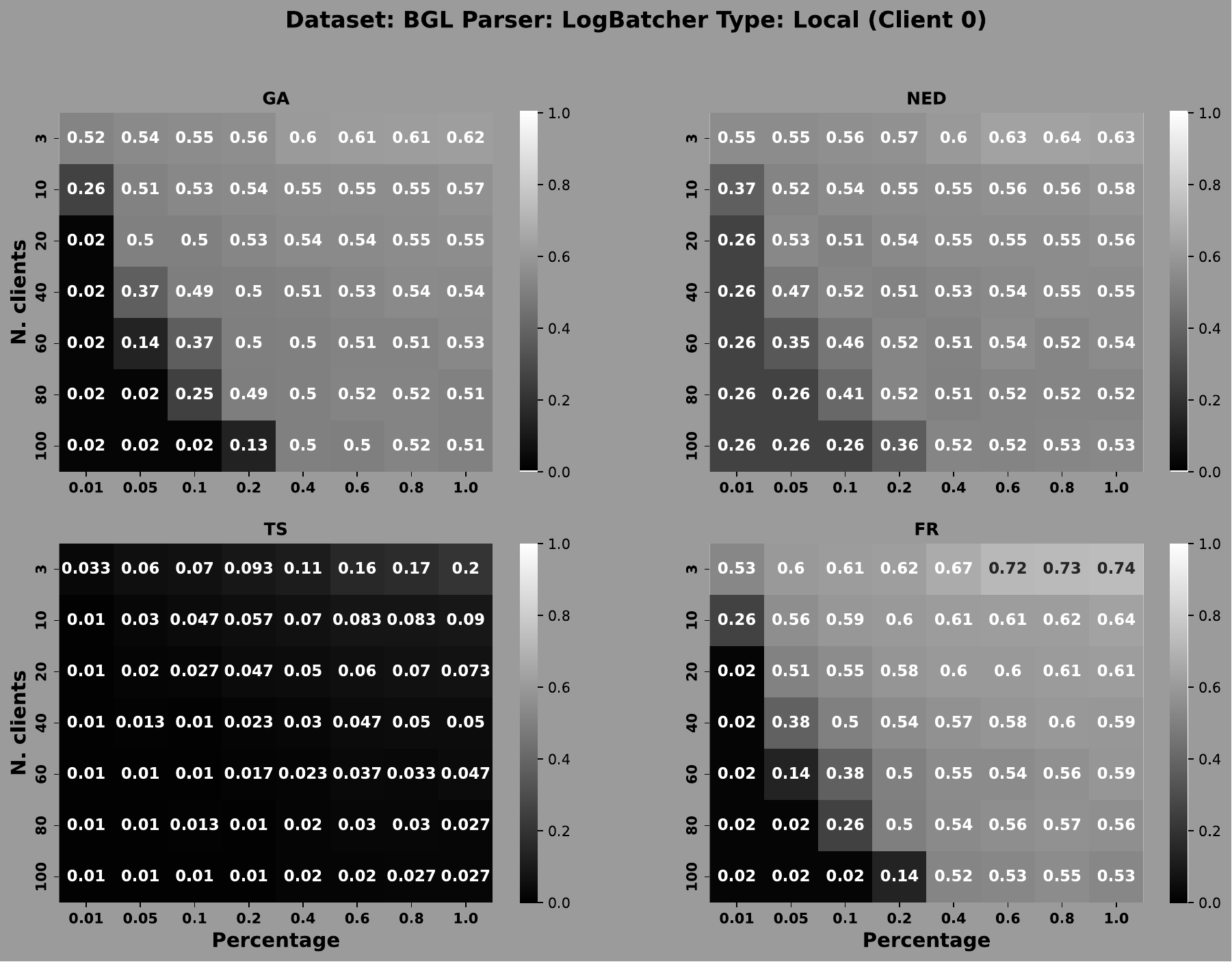}
  \caption{Heatmap illustrating the relationship between the number of clients and the proportion of data in a local configuration for the BGL dataset with LogBatcher.}
  \label{fig:bgl-local_heat_map}
\end{figure*}

\begin{figure*}[t]
  \centering
  \includegraphics[width=0.6\textwidth]{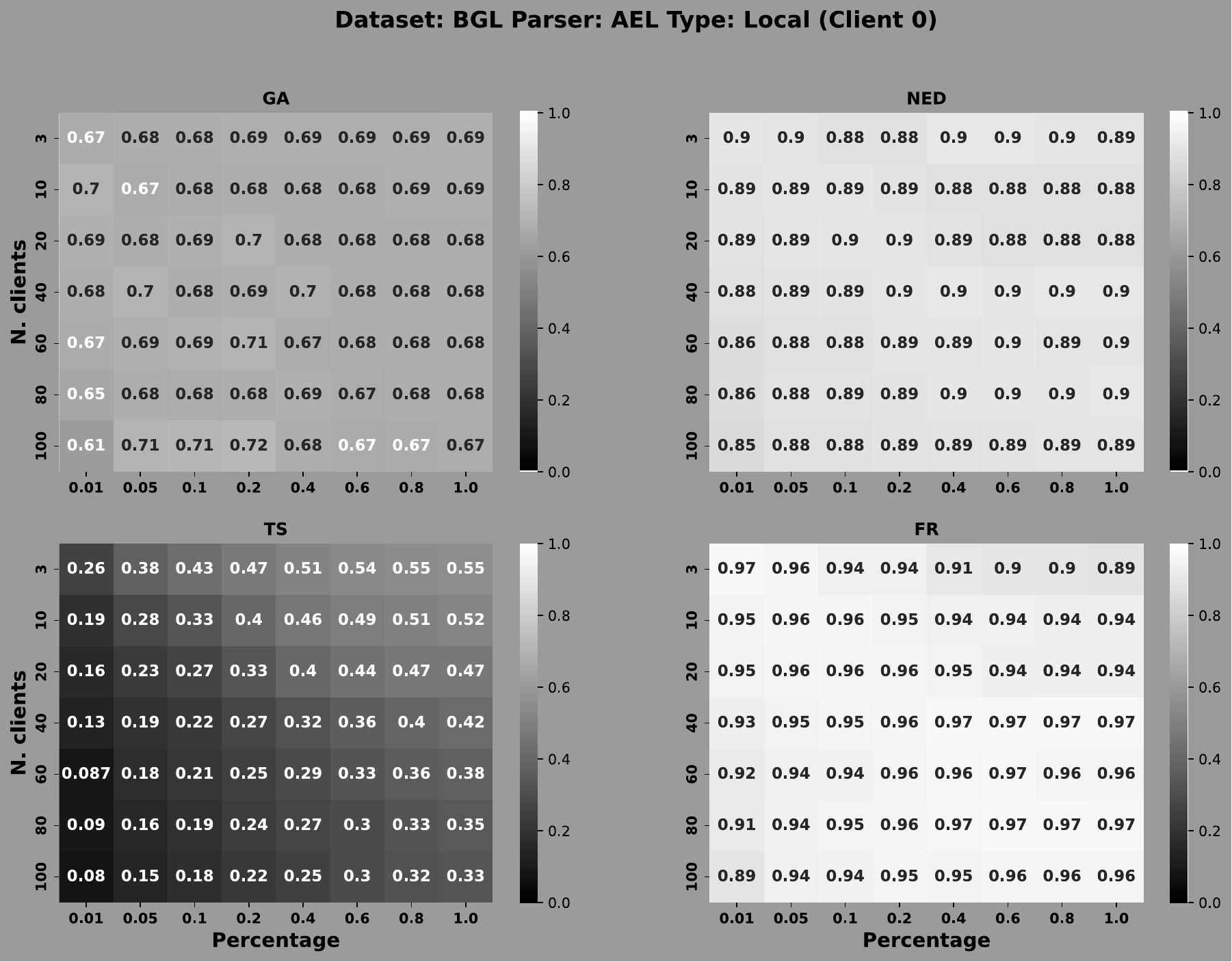}
  \caption{Heatmap illustrating the relationship between the number of clients and the proportion of data in a local configuration for the BGL dataset with AEL.}
  \label{fig:bgl-local_heat_map_ael}
\end{figure*}

A comparable pattern appears in Figures \ref{fig:bgl-local_heat_map} and \ref{fig:bgl-local_heat_map_ael}, where LogBatcher shows greater sensitivity to the data distribution and performs worse on BGL than AEL, a simpler and older method. Additional results for other parsers and metrics are provided in our code repository.

\subsection{Federated setup}

\begin{figure*}[t]
  \centering
  \includegraphics[width=0.6\textwidth]{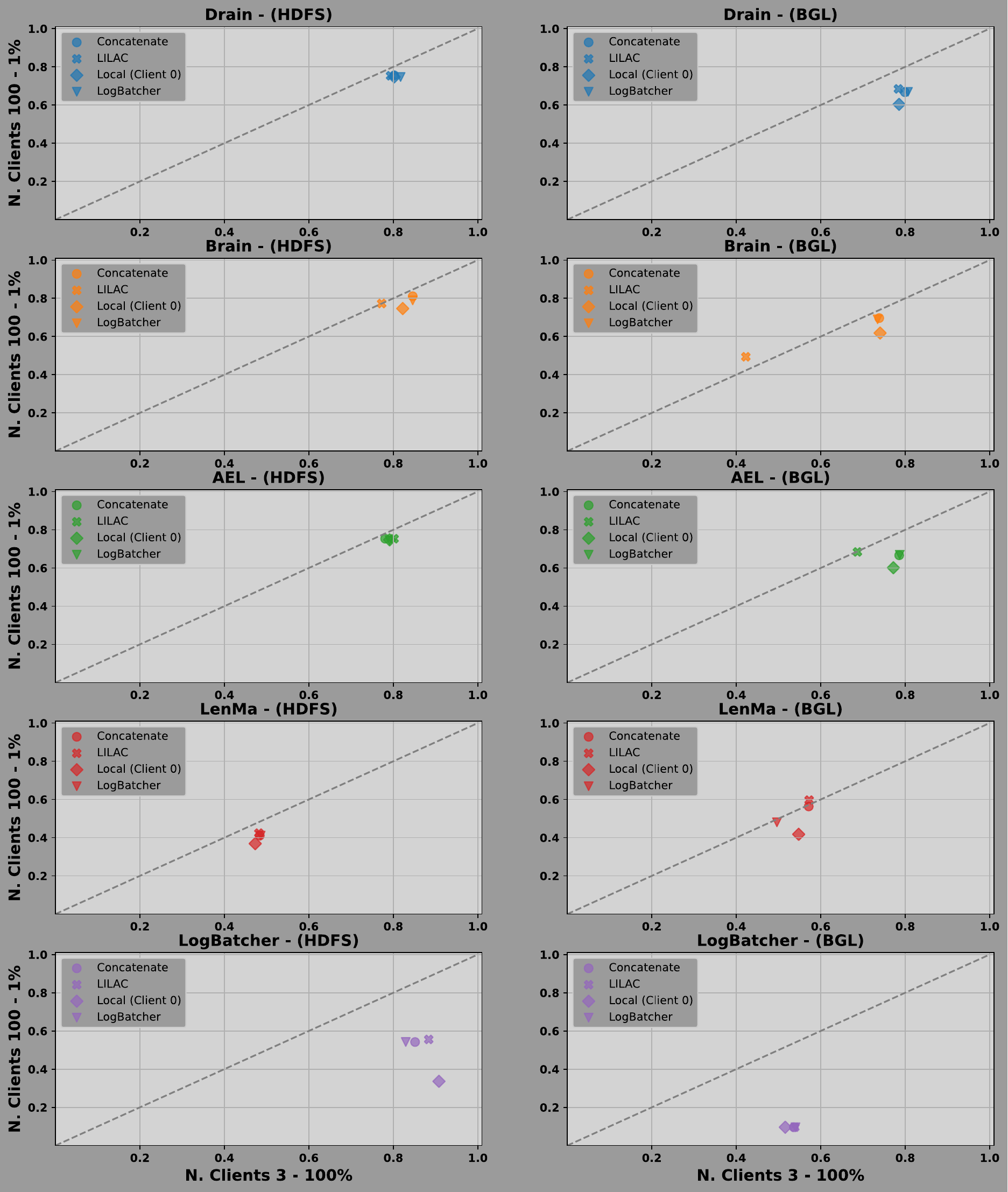}
  \caption{Comparison between different parser setups.}
  \label{fig:hdfs-alpha}
\end{figure*}

\begin{figure*}[t]
  \centering
  \includegraphics[width=0.6\textwidth]{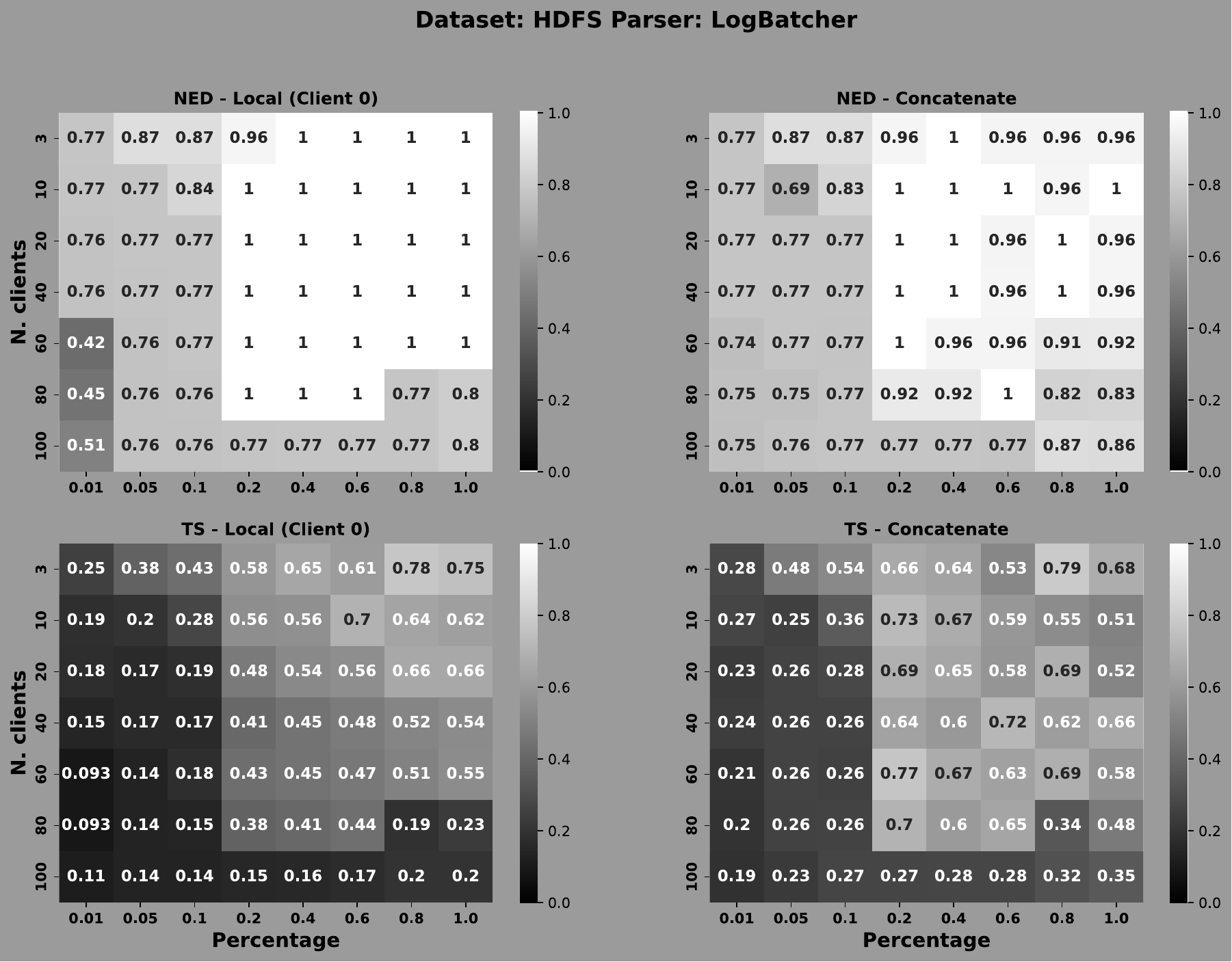}
  \caption{Heatmap illustrating the relationship between the number of clients and the proportion of data in a local and distributed configuration for the HDFS dataset.}
  \label{fig:hdfs-local_heat_map_2}
\end{figure*}

\begin{figure*}[t]
  \centering
  \includegraphics[width=0.6\textwidth]{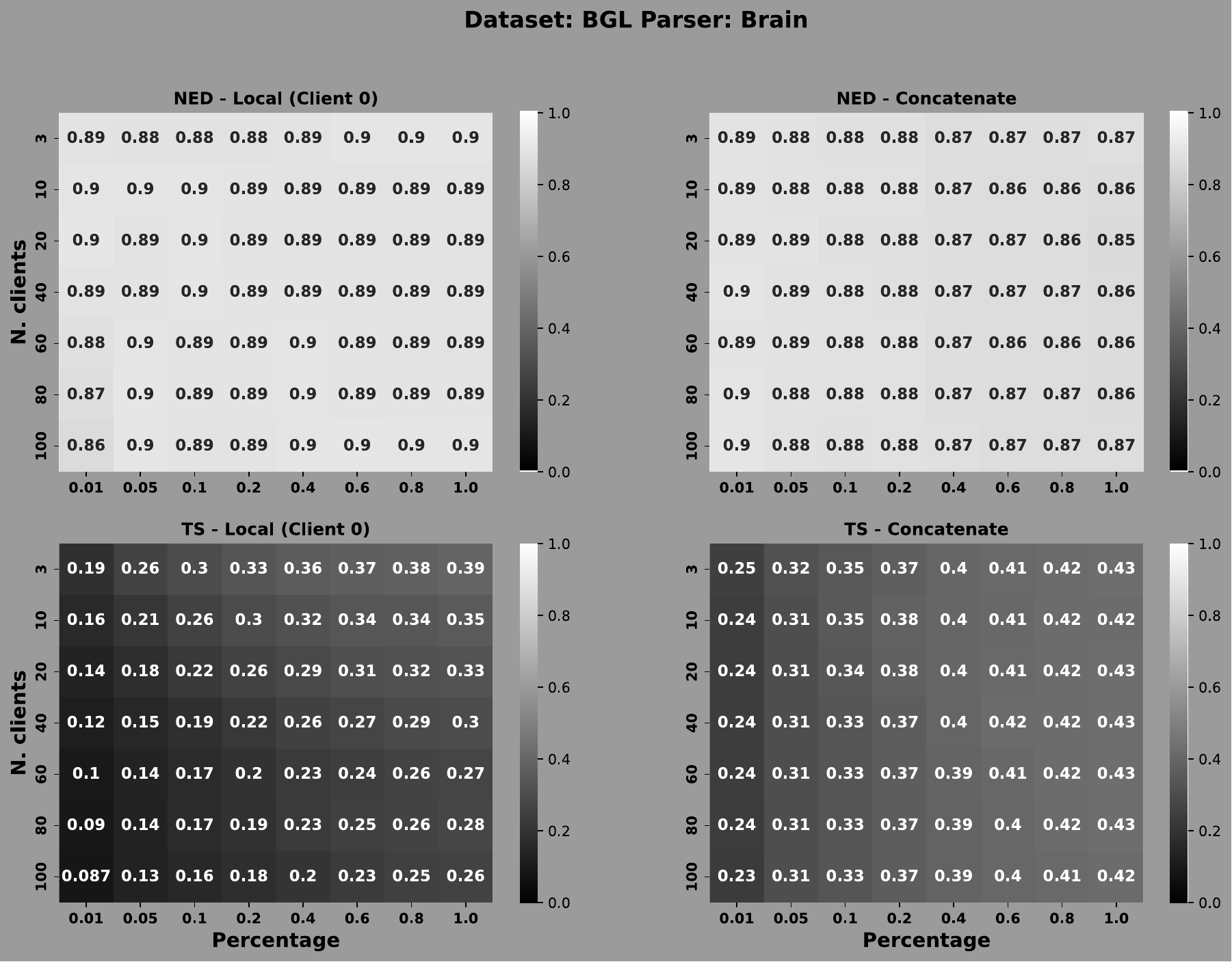}
  \caption{Heatmap illustrating the relationship between the number of clients and the proportion of data in a local and distributed configuration for the BGL dataset.}
  \label{fig:bgl-local_heat_map_2}
\end{figure*}

We contrasted the results from the previous subsection with those generated by the various aggregation methods. Figure \ref{fig:hdfs-alpha} show a comparison, on each axis, between the experiment with 3 clients using 100\% of the data and the experiment with 100 clients using 1\% of the data. We use the metric introduced in Equation \ref{eq:alpha}. An example demonstrating how to interpret this plot is shown in Figure \ref{fig:plot_comparison}. Since this metric combines three different measures into a single representation, we also include Figure \ref{fig:hdfs-local_heat_map_2} and Figure \ref{fig:bgl-local_heat_map_2} to facilitate a comparison between the local client and concatenate methods using NED and FR. A more detailed analysis of the results is provided in Section \ref{sec:disc}. Additional results regarding the application of the aggregation strategies across all configurations can be found in our code repository.

\begin{figure*}[t]
  \centering
  \includegraphics[width=0.75\textwidth]{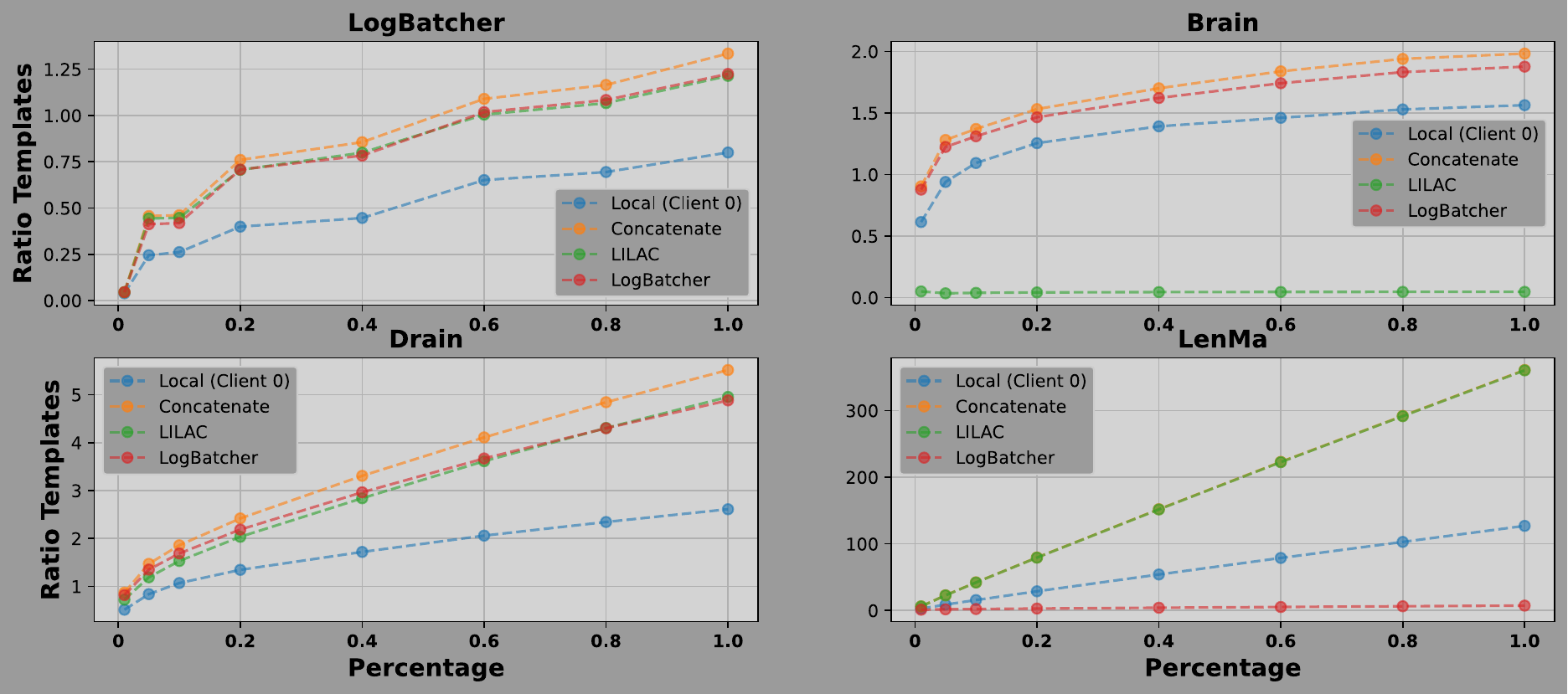}
  \caption{Template ratio comparison between different parsers in BGL with 3 clients setup.}
  \label{fig:bgl_ratio}
\end{figure*}

Figure \ref{fig:hdfs-alpha} indicates that aggregation techniques can generally enhance parser performance, but not in all cases. Overall, BGL yields better performance. This can be understood by looking at Figure \ref{fig:dataset_dist}, which illustrates that BGL contains a broader range of event types. As a result, aggregation becomes especially beneficial in scenarios where there is a large number of log templates distributed across different clients. However, as seen in the local setup, many parsers are not highly sensitive to data distribution. Consequently, techniques such as concatenation increase the number of templates but do not necessarily lead to information gain. In contrast, GA and NED metrics are less affected by aggregation, whereas techniques such as TS are more strongly influenced and exhibit a clear performance improvement. In some situations, such as with LenMA, the number of templates after aggregation grows dramatically, indicating that these methods cannot effectively compress the templates produced by the parsers. Figure \ref{fig:bgl_ratio} illustrates how the template ratio evolves among different parsers on BGL. All aggregation techniques exhibit broadly comparable performance, with only small differences. Nevertheless, since the concatenation approach is the least computationally complex, we consider it a more suitable candidate to be use in future work.

Figures \ref{fig:hdfs-local_heat_map_2} and \ref{fig:bgl-local_heat_map_2} present the results for NED and TS using LogBatcher on HDFS and Brain on BGL, respectively, under both the local and concatenated settings. We can see that LogBatcher, which is more sensitive to data distribution, benefits more from the concatenation-based aggregation, whereas the improvement in NED for Brain is debatable. In both cases, however, TS improves. This suggests that although concatenation helps recover the original templates more accurately, parsers like Brain do not strictly require it, as they can already identify similar templates that fulfill a comparable role.

\section{Result discussion}
\label{sec:disc}

In this section, we provide a more detailed description of the results presented in Section \ref{sec:results}, along with a discussion that addresses all the research questions introduced in Section \ref{sec:introduction} with \textbf{research answers (RA)}.

\textbf{RQ1 How do the number of clients and the volume of log data influence the effectiveness of different parser algorithms?}

\textbf{RA1}: Increasing the number of clients, or alternatively decreasing the amount of data per client, negatively impacts the overall performance of the parser but parsers like LogBatcher are more sensitive to this change. Other parser appears to be more resilient to this problem. The performance degradation is more abrupt for BGL than for HDFS. This can be explained by the event distribution shown in Figure \ref{fig:dataset_dist}: HDFS contains fewer event log types, with a small subset occurring much more frequently than the rest, whereas BGL includes a larger and more evenly distributed set of event types. The experiments show that the logs dataset distribution is the main factor that determines how strongly the different parsers are impacted in a distributed environment. The results further indicate that, for some parsers, using smaller HDFS datasets can actually yield better performance. This is most likely because certain rare log messages can negatively affect the parsers’ overall accuracy. A similar phenomenon was observed by \citet{pustozerovaLightweightTechniquesFederated} in the context of log anomaly detection methods.

\textbf{RQ2 To what extent can different aggregation techniques improve the performance of local parsers in a distributed environment?}

\textbf{RA2}: Aggregation techniques can improve overall performance in distributed settings. However, some benchmark metrics are influenced more strongly than others, suggesting that the impact of aggregating parsing can vary depending on the specific downstream task. Aggregation techniques also appear to improve robustness for datasets such as BGL, especially in LogBatcher, which, as shown in RA1, was more sensitive to the distribution configuration. Other parsers, such as Drain and Brain, yield only limited gains or none at all, particularly on datasets like HDFS.

\textbf{RQ3 What are the main considerations when deploying parsers in distributed environments?}

\textbf{RA3}: While we demonstrate that applying aggregation techniques can enhance overall performance in distributed settings, it also increases system complexity. The agregation step is an additional phase that requires communication between components that would not necessarily need to interact otherwise. This, in turn, introduces a range of privacy risks, such as those highlighted by \citet{pustozerovaLightweightTechniquesFederated}. For example, if a generated template includes a user’s name or a password, it could inadvertently expose personal information to other organizations. To answer this question, we will consider not only the results presented in this paper, but also the broader context of cost–benefit.

One situation in which aggregation parsing is required or strongly recommended is when a downstream component relies on a federated log anomaly detection method. In this case, agregation parsing becomes essential because it aligns event IDs across different templates and maintains consistency between nodes. Another suitable use case is when we are using a LLM-based parser as LogBatcher, as shown in the previous results. On the other hand, if the dataset includes relatively few event types and only a limited number of clients, the marginal performance gain may not be sufficient to offset the added complexity and potential issues introduced by aggregation approaches, making them an impractical choice.

Another aspect to take into account is the communication overhead. If, whenever a new log type is detected, all templates on all nodes must be updated, this can cause a large communication cost, especially during the rollout of a new service or an update that introduces many new log event types in a short period of time. This issue is even more pronounced in a decentralized architecture, where the number of required messages is higher than in a centralized design. On the other hand, although a centralized approach reduces the number of messages, it also introduces a single point of failure into the system. The communication overhead can be reduced by performing the aggregation approach only when each client has discovered a certain number of new templates.

An interesting use case arises when multiple organizations have parsers that have been running long enough to collect sufficient logs and achieve good overall performance. These organizations could share their template sets in a federated way so that, when a new organization or parser comes online that handles a similar software infrastructure, it can immediately leverage those templates. In this manner, new systems can benefit from the expertise and accumulated knowledge of more established systems.

\section{Threat to validity}
\label{sec:validity}

In this study, we conducted several empirical experiments. In this section, we outline the threats we have identified that may jeopardize the validity of our experiments.

\subsection{Ground truth templates}

The ground truth templates are used to evaluate how well the different parsers perform under various configurations. However, these ground truth templates can include minor human mistakes that introduce noise into the final results. Furthermore, previous studies \citep{fu2023empirical, khan2024impact} indicate that, for tasks such as anomaly detection, parsers that produce templates that match the ground truth more closely do not necessarily lead to better detector performance. This raises the risk that the reported results for certain tasks may be misleading.

\subsection{Parser variety}

In this study, we aim to report the results obtained from a large and diverse set of parsers, since relying on a single type of parser could introduce bias into our overall findings. Nevertheless, parser algorithm development is an active and evolving field, and it is possible that the conclusions drawn in this paper may no longer hold true when future parsers are considered.

\subsection{Data leakage}

For our evaluation we utilize the LLM-parser LogBatcher \citep{xiaoDemonstrationFreeMorePractical2024}. Following previous research \citep{beck2025system, jiangLILACLogParsing2024} we note that the pretraining phase of the LLMs could have exposed them to the log data set used in this study, giving them an advantage for log parsing. However, other studies suggest that direct fact memorization is unprobable, unless seen from diverse sources in training \citep{schwarzschildRethinkingLLMMemorization2024} and that LLMs are not able to return the exact logs or templates from a common dataset when asked \citep{beck2025system}.

\section{Conclusion}
\label{sec:conclusion}

In this work, we examine how different parsers operate in a distributed environment. We propose and evaluate several aggregation strategies that can be applied in a federated setting, and we introduce a taxonomy of parsing metrics to more clearly interpret the collected results. In total, we conduct 6,720 distributed parsing experiments and obtain 40,320 individual results, which we analyze to derive key insights and practical recommendations for deploying parsers in distributed environments.

In future work, we will expand the number of datasets by incorporating newly published datasets from the literature \citep{landauer2022maintainable}, and we will develop new aggregation techniques that can be validated against the baselines introduced in this study.


\bibliographystyle{cas-model2-names}
\bibliography{cas-refs}

\end{document}